\documentclass{article}
\usepackage{graphicx}
\usepackage{authblk}
\usepackage{epstopdf, epsfig, amsmath}
\usepackage{caption,color}
\usepackage{subcaption}
\usepackage[letterpaper,margin=1.0in]{geometry}

\title{A model for Faraday pilot-waves over variable topography}

\author[1]{Luiz M. Faria}

\affil[1]{Department of Mathematics, Massachusets Institute of Technology,
Cambridge, MA 02139, USA}

\begin{document}

\maketitle


\begin{abstract}
  Couder and Fort \cite{couder2005dynamical} discovered that droplets walking on a
  vibrating bath possess certain features previously thought to be
  exclusive to quantum systems. These millimetric droplets synchronize
  with their Faraday wavefield, creating a macroscopic pilot-wave
  system.  In this paper we exploit the fact that the waves generated
  are nearly monochromatic and propose a hydrodynamic model capable of
  quantitatively capturing the interaction between bouncing drops and
  a variable topography. We show that our reduced model is able
  to reproduce some important experiments involving the
  drop-topography interaction, such as non-specular reflection and
  single-slit diffraction.
\end{abstract}

\section{Introduction}

The most mysterious manifestation of particle-wave duality certainly
belongs to the quantum world. Our intuition, formed primarily by
experiencing particles and waves at macroscopic scales, appears to be
of little use when trying to rationalize the inherent duality of
subatomic objects. At the heart of quantum mechanics lies this
duality, where electrons are at times waves, diffracting, and at times
particles, hitting a screen.  Recently, a series of fluid dynamics
experiments has provided an interesting example of particle-wave
association at the macroscopic scale. The experiments consist of
releasing a millimetric drop on the surface of a periodically
oscillating bath. Given the right combination of parameters, the drop
avoids coalescence by always maintaining an air layer between itself
and the free-surface \cite{walker1978drops}. Such a drop can bounce
indefinitely on the surface, generating a wave-field around its
position. Furthermore, as the parameter controlling the amplitude of
the periodic forcing is increased, the drops can be shown to go
through an instability and start moving horizontally, giving rise to
what are commonly called \emph{walkers}
\cite{couder2005dynamical}. Since the waves are generated by the
drops, and the drops propelled by the waves, the system comprises a
macroscopic analogy of the pilot-wave theory of quantum mechanics as
first conceived by
\cite{de1930introduction}
.

Several studies have further explored the analogies
between bouncing drops and quantum
mechanics
.
\cite{couder2006single} demonstrated that walkers diffract
through single- and double-slits in a manner reminiscent to
electron/photon diffraction \cite{bach2013controlled}. \cite{eddi2009unpredictable},
showed that walkers can tunnel through submerged boundaries,
analogous to quantum tunneling. \cite{fort2010path} established that,
when in a rotating frame, walkers settle into circular orbits of
quantized radii, providing an analogy to Landau
quantization (see also
\cite{oza2014pilot,harris2014droplets}). \cite{perrard2014self} and
\cite{labousse2014etude} examined walker motion in a central force,
and reported the emergence of a double quantization in energy and
angular momentum. \cite{harris2013wavelike} studied walkers in a circular
bath, and demonstrated the emergence of coherent statistical behavior
similar to what is observed in quantum corrals
\cite{crommie1993confinement}.

Central to the understanding of this hydrodynamic pilot
  wave system is the concept of memory
  \cite{couder2006single,eddi2011information}. Every time the drop
  bounces on the vibrating surface, it generates a wave which decays
  exponentially in time due to the viscous dissipation. Upon several
  impacts, the wavefield is thus the accumulation of all contributions
  from previous impacts. The ability of a drop to ``remember'' its
  past depends crucially on the decay rate of the
  waves
  . Near the Faraday threshold, the waves decay slowly, and the drop's
  dynamics will depend not only on its current position, but also on
  the position of its previous impacts. This is, in essence, the
  source of the temporal non-locality displayed by bouncing drops, and
  represents a key idea in most theoretical models
  \cite{protiere2006particle,couder2006single,eddi2011information,labousse2014etude,molavcek2013dropswalking,oza2013trajectory}.

Despite significant progress in the theoretical understanding of this
hydrodynamic pilot-wave system, an important gap still exists: the
interaction of walkers with variable topography. Such interactions
give rise to apparently non-classical behaviour of the drops and arise
in a number of key hydrodynamic quantum analog systems. Therefore an
ability to model these interactions will be crucial to better
understanding the analogy --and the limitations thereof-- between the
macroscopic pilot-wave system and its quantum counterpart. Most of the
theoretical models developed in the past (see
\cite{bush2015pilot} for a review) focus on baths of
effectively infinite depth. A few recent exceptions include
\cite{nachbin-2016-tunneling}, \cite{gilet2014dynamics},
\cite{blanchette2016modeling}, and \cite{dubertrand2016scattering}. In
\cite{nachbin-2016-tunneling} conformal mapping is used to transform a
variable topography to a flat one, a technique limited to
two-dimensions. In \cite{gilet2014dynamics} and
\cite{blanchette2016modeling} the variation in depth is treated as an
effective boundary, and the wavefield is decomposed into
eigenfunctions of the Laplacian with either a homogenous Neumann
\cite{gilet2014dynamics} or Dirichlet \cite{blanchette2016modeling}
boundary condition. \cite{dubertrand2016scattering} takes an approach
similar to \cite{gilet2014dynamics} in imposing a homogenous Neumann
condition at the effective boundaries, but the equations are solved
employing a Green's function approach instead of an eigenfunction
decomposition as in \cite{gilet2014dynamics}. 

In this paper we propose a reduced model rich enough to capture the
interaction between walkers and variable topography, yet simple enough
to allow for efficient numerical simulation and physical
transparency. Unlike the works cited in the previous paragraph, the
variable depth is treated not as an effective boundary, but as a place
where the wave speed changes, providing a mechanism for wave
reflection. The model exploits the fact that, because the entire
system is driven near resonance, the wavefield is nearly monochromatic
and therefore only certain wavenumbers need to be modeled
correctly. We show that --with a reasonable quantitative agreement--
our model reproduces some important laboratory experiments such as
non-specular reflection \cite{eddi2009unpredictable,pucci2016non} and
the observed pattern of diffraction past a slit reported in \cite{harris2015pilot}.


\section{Model}\label{sec:model}

In this section we present the theoretical model. In \S
\ref{sec:wave-model}, we develop a simplified description of the surface
waves, and in \S \ref{sec:drop-dynamics-wave} we incorporate the effect of
drops bouncing on the surface.

\subsection{Surface waves model}\label{sec:wave-model}

We are interested in the surface waves of a bath undergoing a vertical
sinusoidal oscillation. In the bath's frame of reference, the problem
is equivalent to considering a time dependent gravity:
$g = g_0 (1 + \Gamma \cos{(\Omega t)})$, where $g_0$ is the
gravitational acceleration, $\Omega$ denotes the angular frequency of
oscillation, and $\Gamma$ the amplitude. For a fixed $\Omega$, there
exists a critical amplitude of oscillation, $\Gamma_F$, above which
the surface becomes unstable. In the nearly inviscid limit of interest
to this paper, the instability is always subharmonic, with frequency
$\omega_0 = \Omega / 2$
\cite{benjamin1954stability,kumar1996linear}. The bouncing drop
experiments described in the introduction are typically carried out
near (and below) the stability boundary, so that the flat surface of
the bath is stable, but subharmonic modes excited within it decay very
slowly. Furthermore, the drops are small enough that the waves they
generate may be treated as linear.

Instead of beginning with the linearized Navier-Stokes equations, we
take as a starting point the quasi-potential theory developed in
\cite{milewski2015faraday}, which when extended to a bath of finite
depth is given by
\begin{align}
  \label{eq:PM-1}
  \Delta \phi  &= 0, && \text{for }-h(\mathbf{x})\leq z\leq 0\\
  \label{eq:PM-2}
  \nabla \phi \cdot \mathbf{n} &= 0, &&\text{for } z=-h(\mathbf{x})\\
  \label{eq:PM-3}
  \phi_t &= -g(t) \eta + 2\nu^* \Delta_\perp \phi+\frac{\sigma}{\rho} \Delta_\perp
           \eta, &&\text{for } z=0\\
  \label{eq:PM-4}
  \eta_t &= \phi_z + 2\nu^* \Delta_\perp \eta, &&\text{for } z=0
\end{align}
Here $\eta, \phi$, denote the free-surface displacement and velocity
potential. The parameter $\nu^*$ is an
effective kinematic viscosity, chosen to capture the correct stability
threshold $\Gamma_F$, and $\sigma, \rho$ denote
surface tension and density. Finally, $\mathbf{n}$ is a vector normal
to the bottom profile, $h(\mathbf{x})$ the depth of the undisturbed
surface, and $\Delta_\perp$ the horizontal Laplacian (i.e.
$\Delta_\perp = \partial_{xx} + \partial_{yy}$).

We consider a bath having a deep and
a shallow region, separated by a discontinuous jump, so
$h(\mathbf{x})$ is given by:
\begin{equation}
  \label{eq:depth}
  h(\mathbf{x}) = 
  \begin{cases}
    h_0 \qquad \text{for } \mathbf{x} \in \mathcal{D},\\
        h_1 \qquad \text{for } \mathbf{x} \not \in \mathcal{D},
    \end{cases}
\end{equation}
where $h_0>h_1$. The key feature that we aim to exploit is that the
sinusoidal oscillations of the bath favor the resonating temporal
frequencies, causing the waves to be nearly monochromatic. Thus we
need not model all waves correctly, but only
those for which the frequency is approximately $\Omega/2$; other waves
are less important owing to their faster temporal decay. 

The main difficulty in solving \eqref{eq:PM-1}--\eqref{eq:PM-4} is
that $\phi_z$ in equation \eqref{eq:PM-4} couples the surface
evolution to Laplace's equation in the three-dimensional domain,
making the problem computationally expensive. When
$h(\mathbf{x}) = \text{constant}$, it can be shown that (see
e.g. \cite{whitham1974linear})
\begin{equation}
  \label{eq:DtN-flat-bottom}
  \mathcal{F}(\phi_z(\mathbf{x},0,t)) =
  |\mathbf{k}|\tanh{(|\mathbf{k}| h)}
  \mathcal{F}(\phi(\mathbf{x},0,t)), 
\end{equation}
where $\mathcal{F}$ denotes the two-dimensional Fourier transform in
$x$ and $y$, and $\mathbf{k} = (k_x, k_y)$ is the wavenumber
vector. This allows for the surface evolution to be represented, in
Fourier space, using only data on the surface, reducing a 3
dimensional problem to 2 dimensions. When $h(\mathbf{x})$ is not
constant, however, no simple explicit representation of
$\phi_z(\mathbf{x},0,t)$ in terms of $\phi(\mathbf{x},0,t)$ exists
.
Our goal is to approximate $\phi_z$ in equation \eqref{eq:PM-4} using
data on the surface only, and to do so in a way that correctly models waves with frequency $\omega = \Omega/2$.

Motivated by shallow water theory, we seek an approximation of the
form
\begin{align}
  \label{eq:wave-eq-approximation}
\phi_z(\mathbf{x}, 0, t) \approx -\nabla_\perp \cdot (b(\mathbf{x})
\nabla_\perp \phi(\mathbf{x},0,t))  
\end{align}
and choose $b(\mathbf{x})$ to exactly match the dispersion
relation of Faraday waves in both the shallow and deep regions. Given
$\Omega$ and $h_0,h_1$, we must thus determine the most unstable
wavenumber in each region, denoted by $k_{F_0},k_{F_1}$. 
Following Appendix A of \cite{milewski2015faraday}, the Faraday
wavenumber in each region of constant depth is computed by finding the $k$ which minimizes $\Gamma$ in the
following expression:
\begin{align}
  \label{eq:neutral-stability-relation-quasi-potential}
\left[\omega_I^2 + \gamma^2 - \frac{1}{4}\Omega^2  \right]^2 +
  \Omega^2 \gamma^2 - \frac{1}{4}{\omega_g^2\Gamma^2} = 0,
\end{align}
where $\omega_I^2(k) = k \tanh{(k h)} (g_0 + \sigma/\rho k^2)$ is the
ideal/inviscid dispersion relation, $\gamma = 2\nu^*k^2$ is the dissipation
rate, and $\omega_g = g_0k\tanh{(kh)}$. Notice that this reduces to the well-known result of
\cite{benjamin1954stability} when $\gamma\to 0$, and in that limit
$k_{F_0}, k_{F_1}$ simply satisfy
\begin{align}
  \label{eq:inviscid-dispersion-relation}
\left(g_0
  + \frac{\sigma}{\rho} k_{F_0}^2 \right) k_{F_0} \tanh{(k_{F_0} h_0)} = \left(g_0
  + \frac{\sigma}{\rho} k_{F_1}^2 \right) k_{F_1} \tanh{(k_{F_1} h_1)} = \Omega/2.
\end{align}

In order for \eqref{eq:wave-eq-approximation} to exactly match the
dispersion relation of Faraday waves in
the regions of constant depth, we must choose
\begin{equation}
  \label{eq:effective-depth}
  b(\mathbf{x}) = 
  \begin{cases}
    \tanh{(k_{F_0}h_0)} / k_{F_0} \qquad \text{for } \mathbf{x} \in \mathcal{D},\\
       \tanh{(k_{F_1}h_1)} / k_{F_1}\qquad \text{for } \mathbf{x} \not \in \mathcal{D}.
    \end{cases}
  \end{equation}
This leads to the following wave model:
\begin{align}
  \label{eq:wave-model-variable-depth-1}
  \phi_t &= -g(t) \eta + 2\nu^* \Delta_\perp \phi+\frac{\sigma}{\rho} \Delta_\perp
           \eta,\\
  \label{eq:wave-model-variable-depth-2}
  \eta_t &= -\nabla_\perp \cdot (b(\mathbf{x}) \nabla_\perp \phi
           )+ 2\nu^* \Delta_\perp \eta, 
\end{align}
where $b$ is given by equation \eqref{eq:effective-depth}.

It is worth noting that in the long wave limit (i.e. $k h \to 0$),
we have that $b = h$, and formally the linear shallow water
theory is recovered. Our approximation, however, does not rely on
$k_{F_0} h_0 \ll 1$, which is not typically true for the bouncing
drop experiments, but on the assumption that the wave-field is nearly
monochromatic, with all the energy in modes of temporal
frequency near $\Omega/2$. For drops bouncing periodically over a vibrating
bath, parametric resonance ensures that such an approximation is
justifiable, as later shown in the quantitative comparison performed
in Figure \ref{fig:wavefield-comparison}. In the next section we
modify the wave model given by equations
\eqref{eq:wave-model-variable-depth-1}--\eqref{eq:wave-model-variable-depth-2}
in order to account for a drop bouncing on its surface. 



\subsection{Drop dynamics and wave/drop coupling}\label{sec:drop-dynamics-wave}

The wave-drop coupling is identical to that of 
\cite{milewski2015faraday} and \cite{molavcek2013dropswalking}, where the drop is
treated as an excess pressure on the dynamic surface condition. This
leads to the following system:
\begin{align}
  \label{eq:governing-equations-dimensional-1}
  \phi_t &= -g(t) \eta + \frac{\sigma}{\rho} \Delta_\perp \eta + 2\nu^* \Delta_\perp \phi-
             \frac{1}{\rho} P_D(\textbf{x}-\textbf{x}_p(t),t)\\
  \label{eq:governing-equations-dimensional-2}
  \eta_t &= - \nabla_\perp \cdot (b \nabla_\perp \phi) + 2\nu^* \Delta_\perp \eta.
\end{align}
The new term $P_D$ represents the effect of the drop, and
$\mathbf{x}_p$ denotes the drop's horizontal position, which
evolves according to 
\begin{equation}
  \label{eq:EOM-horizontal-dimensional}
  m \frac{d^2\textbf{x}_p}{dt^2} + \left(c_4 \sqrt{\frac{\rho
        R_0}{\sigma}} F(t) + 6 \pi R_0 \mu_{air}\right)
  \frac{d\textbf{x}_p}{dt} = -F(t) \nabla \eta|_{\textbf{x} = \textbf{x}_p}.
\end{equation}
The parameters $m, R_0, \mu_{\text{air}}$, and $c_4$ denote the
drop mass, drop radius, air viscosity, and coefficient of
tangential restitution, respectively \cite{molavcek2013dropswalking}. The function $F(t)$ represents
the reaction force exerted on the drop by the fluid.

Calculating $F(t)$ requires solving for the vertical dynamics of the
particle, which as shown in \cite{molavcek2013dropsbouncing} can be
well approximated by a logarithmic spring during contact. Although it
is possible to couple the wave model here presented to the log-spring
model of \cite{molavcek2013dropsbouncing}, for the sake of simplicity
we assume that the drop bounces periodically, with period
$T_F = 4\pi/\Omega$. It can then be shown (see
\cite{molavcek2013dropswalking}) that
$\int_t^{t+T_F} F(\tau) d\tau = mg_0T_F$. Assuming that the contact
time is small enough (relative to the bouncing period) that $F$ may be
treated as an impulse force, and taking the impacts to happen at
$t=nT_F$, we obtain
$F(t) = mg_0 \sum_{n=0}^\infty\delta\left( \frac{t -
    nT_F}{T_F}\right)$,
where $\delta$ is the Dirac delta function. This eliminates the need
to solve for the vertical motion of the drop;  however, by
  considering the impact to be instantaneous, the model ignores the
  evolution of the waves during contact, which may be important for
  certain types of interaction given that the contact time can be as
  large as one-fourth of the period.  Finally the excess
pressure $P_D$ may be modeled as
$ P_D(\mathbf{x}-\mathbf{x}_p,t) = F(t)
I(||\mathbf{x}-\mathbf{x}_p||)$,
where $I$ describes the spatial distribution of the excess pressure
( and has units of $1/\text{length}^2$).

It is now convenient to nondimensionalize the equations using a
representative length scale, $\lambda_F$, and time scale
$T_F$. 
Using tildes for dimensionless variables, we obtain
\begin{align}
\label{eq:governing-equations-dimensionless-1}
  \tilde \phi_{\tilde  t} &= -G(1 + \Gamma \cos(4\pi \tilde  t - \varphi))
                              \tilde \eta + \frac{2}{\mathrm{Re}}
                              \Delta \tilde \phi+\mathrm{Bo} \Delta
                              \tilde \eta -M G \sum_{n=0}^\infty\delta(\tilde  t-n) \tilde{I}(||\tilde
                              {\textbf{x}}-\tilde{\textbf{x}}_p||)\\
  \label{eq:governing-equations-dimensionless-2}
\tilde\eta_{\tilde  t} &= -\nabla \cdot (\tilde  b \nabla \tilde \phi) + \frac{2}{\mathrm{Re}}\Delta \tilde \eta
\end{align}
where $G = g_0 T_F^2/\lambda_F$,  $\mathrm{Re} = \lambda_F^2/(t_0\nu)$,
$\mathrm{Bo} = \sigma T_f^2 / (\rho \lambda_F ^3)$, $ M=m/ (\rho
\lambda_F^3)$, $\tilde b = b/\lambda_F,$
and $\varphi$ denotes the impact phase.  Similarly equation
\eqref{eq:EOM-horizontal-dimensional} becomes
\begin{equation} \label{eq:EOM-horizontal-dimensionless}
  \frac{d^2\tilde{\textbf{x}}_p}{d\tilde t^2} +
  \left(C_{i} \sum_{n=0}^{\infty}\delta(\tilde t-n) + C_{\text{air}}\right)
  \frac{d\tilde{\textbf{x}}_p}{d\tilde t} = -G
  \sum_{n=0}^\infty\delta(\tilde t-n)
  \nabla \tilde \eta|_{\tilde{\textbf{x}} = \tilde{\textbf{x}}_p}
\end{equation}
where $C_i = c_4 \sqrt{\rho R_0/\sigma}gT_F$ and $C_\text{air} = 6 \pi R_0 \mu_\text{air}T_F/m$.

The last simplification comes from considering a point
impact:
$\tilde I(\tilde{\mathbf{x}}-\tilde{\mathbf{x}}_p) =
\delta(\tilde{\mathbf{x}}-\tilde{\mathbf{x}}_p)$. Equations
\eqref{eq:governing-equations-dimensionless-1}--\eqref{eq:EOM-horizontal-dimensionless}
represent the final form of our model, and the rest of this paper is devoted to their
study.  Henceforth, we drop the tilde notation, but all variables are
assumed dimensionless unless otherwise stated.


\section{Numerical results}\label{sec:numer-simul-comp}

We solve equations
\eqref{eq:governing-equations-dimensionless-1}--\eqref{eq:EOM-horizontal-dimensionless}
numerically using a pseudo-spectral method in space, and fourth order
Runge-Kutta scheme for the time integration. Because of the temporal
delta functions, we resolve the drop/surface interaction analytically
during impact. That is, at $t_i = {1,2,...}$ we modify $\phi$ and
$\mathbf{x}_p$ by integrating our equations from immediately before
until immediately after the impact (see Appendix
\ref{sec:deta-numer-algor} for details on the numerical algorithm). 
Doing so allows for high order accuracy in time without having to
resolve the details of the drop/bath interaction. Convergence tests
performed indicate that indeed fourth-order accuracy is achieved, and
that the results do not depend substantially on the number of spectral
modes employed (Appendix \ref{sec:conv-tests-numer}). 

For the numerical simulations that follow, we choose parameters
appropriate for a fluid of $20$ cSt shaking at a frequency of $80$
Hz. Surface tension is $0.0206$ N/m, and the density is $949$
kg/$\text{m}^3$. In the deep region the depth is $h_0=6.09$ mm. When
obstacles are present the depth above them is (unless otherwise indicated) $h_1 = 0.42$ mm, chosen
to correspond roughly to the experiments of \cite{pucci2016non} and
\cite{harris2015pilot}. Computing $k_F$ using the method outlined in
\S \ref{sec:wave-model} yields $k_{F_0} \approx 1.265\text{mm}^{-1}$ mm and
$k_{F_1} \approx 1.555\text{mm}^{-1}$; $b(\mathbf{x})$ is then computed using
\eqref{eq:effective-depth}. In order to compare our simulations to
available experiments, two different drop sizes are considered: in
subsections \S~\ref{sec:freespace}--\ref{sec:refl-from-plan} the drops
have a radius of $0.39$mm, while in \S~\ref{sec:diffr-past-slit} they
have a radius of $0.335$mm.  The coefficient of tangential restitution
$c_4$ is taken to be $0.17$, as suggested in
\cite{molavcek2013dropswalking} for a bath of $20$ cSt shaking at $80$
Hz. Finally, the effective viscosity is $\nu^* = 0.8025 \nu$, which
yields $\Gamma_F = 4.22$ as the Faraday threshold for the fluid
considered. The computations were performed on a domain of
(dimensionless) size $64\times 64$ using $512 \times 512$ points,
which was sufficient to guarantee resolution-independent results
(see Appendix \ref{sec:conv-tests-numer}). All parameters used are summarized in Table
\ref{tab:physical-parameters}.

\begin{table}
  \centering
  \begin{tabular}{ l | l | l }
    \hline
    Symbol \quad & Meaning & Typical value \\
    $\nu$ & Dynamic viscosity & 20 cSt \\
    $\nu^*$ & Effective viscosity & $0.8025\nu = 16.05$ cSt\\
    $\sigma$ & Surface tension & 0.0206 N/m \\
    $\rho$ & Density & 949 kg/$\text{m}^3$ \\
    $\Omega$ & Driving frequency & 80 Hz \\
    $\Gamma$ & Driving amplitude  & 0 -- 4.22 \\
    $\Gamma_F$ & Faraday threshold for $\Gamma$ & 4.22 \\
    $h_0$ & Depth in the deep region& 6.09 mm \\
    $h_1$ & Depth in the shallow region & 0.42 mm \\
    $k_{F_0}$ & Faraday wavenumber in the deep region & 1.265 $\text{mm}^{-1}$\\
    $k_{F_1}$ & Faraday wavenumber in the shallow region & 1.555 $\text{mm}^{-1}$\\
    $c_4$ & Coefficient of tangential restitution & 0.17 \\
    \hline
  \end{tabular}
  \caption{Physical parameters used in the simulations}
  \label{tab:physical-parameters}
\end{table}

We present below
three examples that illustrate the capabilities of the model. First,
in \S~\ref{sec:freespace}, we briefly study the dynamics of a drop in
a bath of constant depth. We then explore in
\S~\ref{sec:refl-from-plan} the interaction of walkers with a planar
submerged step, demonstrating that our model captures many of the
important features of walker reflection. Finally, we present in \S
\ref{sec:diffr-past-slit} results of diffraction past a single slit.

\subsection{Constant depth }\label{sec:freespace}

We consider first walkers in a uniform bath as a benchmark for the
model. As expected, in the absence of barriers the drops move at a
constant speed that is independent of the initial conditions. For any
fixed set of physical parameters, the final free-space velocity will
depend only on the impact phase $\varphi$. Because we have chosen not
to consider the details of the vertical dynamics, $\varphi$ is extraneous
to the model and may therefore be treated as a free parameter. As in
previous strobed models that average out the vertical dynamics
(e.g. \cite{oza2013trajectory}, \cite{labousse2014etude}), $\varphi$
is chosen so that the theoretical walking speed approximately agrees with the
experimentally reported value.  A summary of the average experimental walking
speed, as well as the inferred impact phase used for all configurations
reported in this paper can be found in Table \ref{tab:phase}.

%
\begin{table}
  \centering
  \begin{tabular}{ c | c | c | c | c | c }
    \multicolumn{4}{c}{Drop radius 0.39mm} &
                                             \multicolumn{2}{|c}{Drop radius 0.335mm} \\
    \hline
    \multicolumn{2}{c|}{$\Gamma/\Gamma_F =0.9$} &
\multicolumn{2}{c|}{$\Gamma/\Gamma_F =0.99$}  &  
\multicolumn{2}{c}{$\Gamma/\Gamma_F =0.99$} \\
    \hline
    v(mm/s) & $\varphi/2\pi$& v(mm/s) & $\varphi/2\pi$& v(mm/s) & $\varphi/2\pi$\\
    11.1 & 0.345 & 11.45 &0.3 &6.6&0.275   
  \end{tabular}
  \caption{Experimentally measured speed, $v$, at different memories for the
  two drop sizes considered in this paper. The phase $\varphi$ is
  obtained by requiring the numerically observed speed to match the
  experimental value, and is in the range consisten with experimental
  observation \cite{molavcek2013dropsbouncing}. }
  \label{tab:phase}
\end{table}

\begin{figure}
    \centering
      \includegraphics[width=1\textwidth]{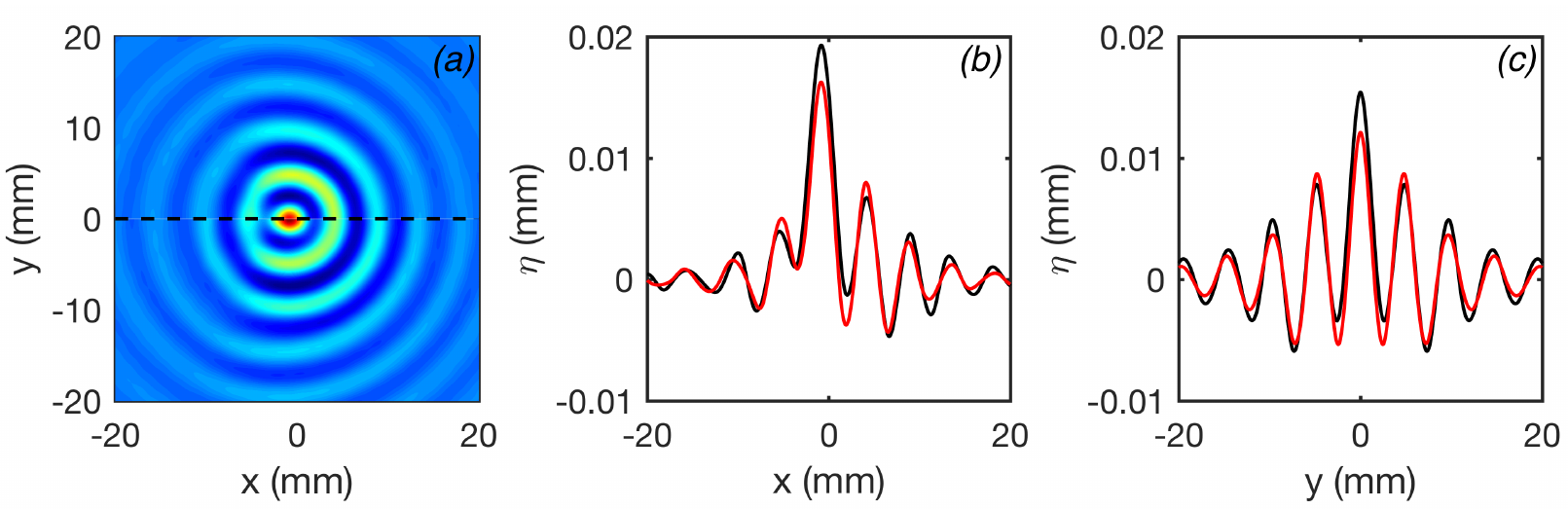}
      \caption{(a) Comparison between our approximation ($y>0$),
       and the quasi-potential theory ($y<0$) of
       \cite{milewski2015faraday} for a
        walker using exactly the same
        parameters. Figures (b) and (c) show the x and y cross
        sections of the wave field. The red line
        represents the quasi-potential theory, and the black line our
        model. Here $R=0.38$mm,
        $\Gamma/\Gamma_F = 0.9$, and the drop is located at
        the origin.}
\label{fig:wavefield-comparison}
\end{figure}

The first comparison we perform is between our model and the equations
they are intended to approximate,
specifically \eqref{eq:PM-1}--\eqref{eq:PM-4}. We show in Figure
\ref{fig:wavefield-comparison} the wave-field generated by a walker
using our nearly monochromatic approximation, and compare it to the
wave-field obtained when using the quasi-potential theory of
\cite{milewski2015faraday}. Although the match is not perfect, the
overall wave features appear to be correctly captured. This indicates
that, at least in places of constant depth, where it is easy to solve for
$\phi_z(\mathbf{x},0,t)$ exactly using \eqref{eq:DtN-flat-bottom},
our monochromatic approximation
$\phi_z(\mathbf{x},0,t)\approx \nabla \cdot ( b \nabla
\phi(\mathbf{x},0,t))$
performs reasonably well. In the next sections, when considering a
variable bottom topography, it becomes very difficult to solve
\eqref{eq:PM-1}--\eqref{eq:PM-4}; therefore, in \S
\ref{sec:refl-from-plan} we instead compare our model to experimental
observations of \cite{pucci2016non} and \cite{harris2015pilot}.

\subsection{Reflection from planar boundary}\label{sec:refl-from-plan}

\begin{figure}
    \centering
    \begin{subfigure}[h]{1\textwidth}
      \includegraphics[width=\textwidth]{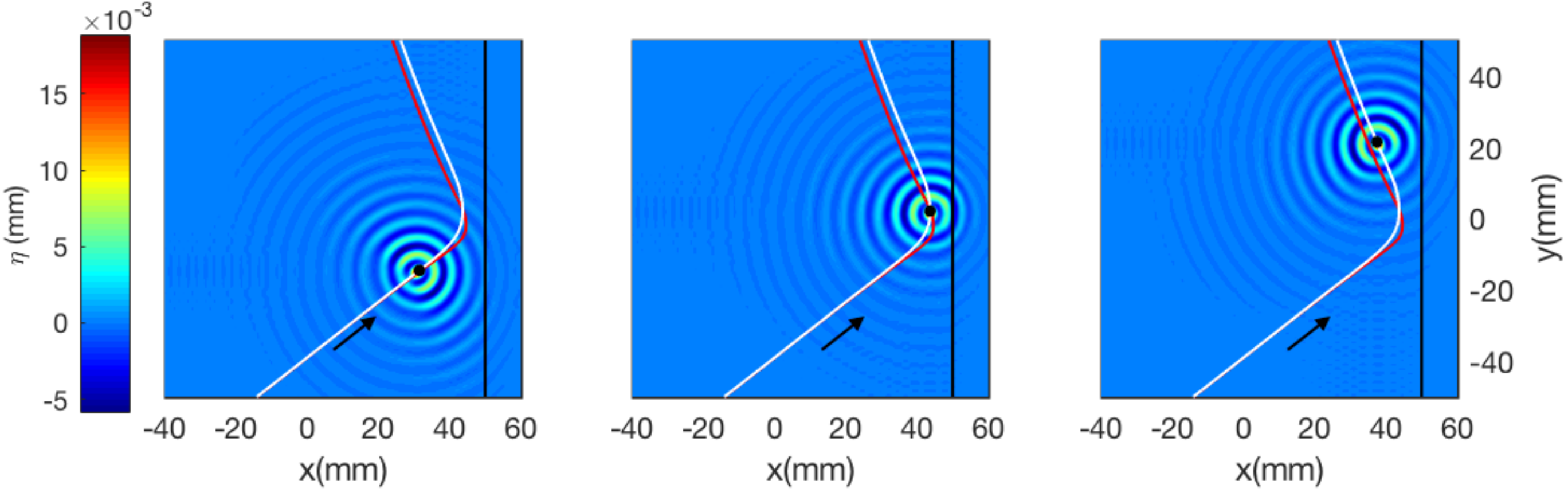}
        \caption{$\Gamma/\Gamma_F = 0.90$. }
        \label{fig:wall-reflection-snapshots-90-mem}
    \end{subfigure}
    ~ 
    \begin{subfigure}[h]{1\textwidth}
        \includegraphics[width=\textwidth]{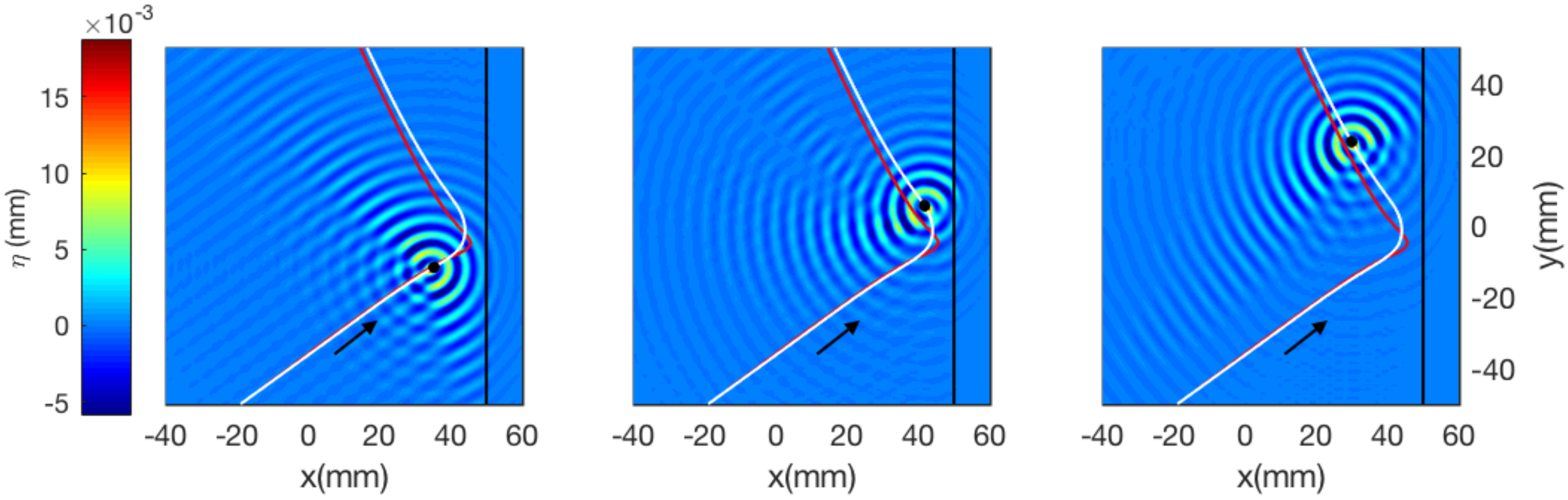}
        \caption{ $\Gamma/\Gamma_F = 0.99$.}
      \label{fig:wall-reflection-snapshots-99-mem}
    \end{subfigure}
    ~ 
    \caption{Wave amplitude (contour plot) and trajectories (white
      line) for wall reflection. The black line at $x=50$ divides the
      deep ($x<50$) and shallow ($x>50$) regions. The leftmost and
      rightmost images are taken 60 bounces before and after the turning
      point, respectively. The center images depict the wave-field exactly at the
      turning point. The red line represents the corresponding
      experimental data \cite{pucci2016non}.}
\label{fig:wall-reflection-snapshots}
\end{figure}

We consider now one of the simplest types of walker-boundary
interactions: reflection from a planar submerged obstacles, and
compare two of our trajectories to the experimental data of
\cite{pucci2016non}. The simulation consists of sending a walker towards
a shallower region at a given incident angle and then
recording both the trajectory and the wave-field for several hundred
impacts. The walker-boundary interaction can be divided into three
distinct stages, depicted in Figure
\ref{fig:wall-reflection-snapshots} for two different values of
$\Gamma/\Gamma_F$. First, at sufficient distance from the shallow
region ($x>50$mm in Figure \ref{fig:wall-reflection-snapshots}), the walker does not feel the presence of the step, and
therefore moves in a straight line. Then, as it approaches the
barrier, its reflected waves start to affect the dynamics, eventually
causing the walker to turn around. The drop continues to interact with
the submerged step through its wave-field as it moves away from it:
the trajectory appears to curve towards the step, becoming asymptotic
to a rectilinear motion.
At high memory (Figure \ref{fig:wall-reflection-snapshots-99-mem}), we
also see the formation of interference fringes caused by the reflected
waves. Finally, we observe that although the model approximately
captures the correct reflection angle (experimental and numerical
lines are nearly parallel in Figure
\ref{fig:wall-reflection-snapshots}), the details of the trajectory
near the wall are slightly different from the experiments, where the
experimental trajectories turn more sharply. Interestingly, the
reflection is non-specular (i.e. incident and reflected angles are not
equal), and there is a small preferred range of reflected angles. A
detailed study of the walker's non-specular reflection, as well as a
more through comparison between the theory here developed and experiments,
is reported in \cite{pucci2016non}. We also see in Figure \ref{fig:wall-reflection-snapshots} that, as
expected, the waves have a much smaller amplitude in the shallow
region ($x>50$) than in the deep region ($x<50$). However, neither
$\eta_x$ nor $\eta$ appears to be zero at the deep-shallow
interface. This inference, consistent with the schlieren imaging of
\cite{eddi2009unpredictable} and \cite{damiano2016surface}, suggests
that imposing an effective Dirichlet or Neumann boundary condition on
the wavefield at places where the depth changes is likely to be
inadequate.

\begin{figure}
    \centering
      \includegraphics[width=\textwidth]{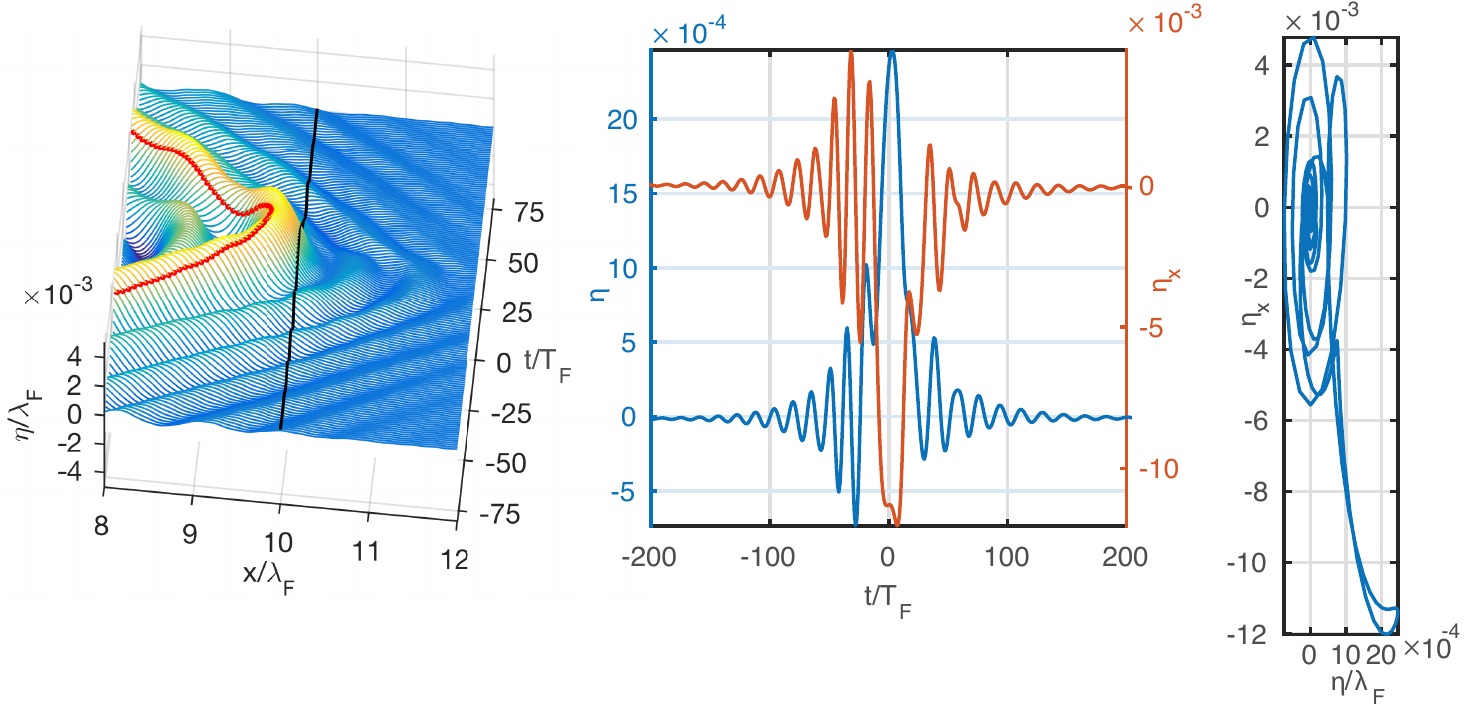}
      \caption{Behavior of the wavefield at the interface between the
        shallow (0.42mm) and the deep (6.09mm) regions.
        In a) we plot the $x$ cross-section of the wavefield of a walker impinging
        perpendicularly upon a step, as a function of time. The
        walker's trajectory is given by $(x(t),0)$), 
        and is shown in red. In b) we plot both $\eta$ and its normal
        derivative, $\eta_x$. In c) we show the relation between
        $\eta$ and $\eta_x$ at $(x,y) = (10,0)$ through the reflection
        process.}
\label{fig:wavefield-above-step}
\end{figure}

In order to study in more detail what happens to $\eta$ at
  the deep-shallow interface, we consider in Figure
  \ref{fig:wavefield-above-step} a walker impinging perpendicularly
  upon the step. The drop's trajectory remains orthogonal to the step
  at all times, allowing for an easier study of the waves by
  focusing on the x cross-section of the wavefield. We show
  in Figure \ref{fig:wavefield-above-step}a the x cross-section of the wavefield as a function of
  time (i.e. $\eta(x,0,t)$), where $t=0$ has been chosen as the time
  where the drop turns away from the shallow region. We see that
  within about $1.5$ wavelengths into the shallow region the wave
  amplitude has already decayed to nearly zero. The behavior of $\eta$ above
  the step, indicated by the black line in the figure, suggests that
  neither $\eta=0$ nor $\eta_x=0$ holds at the deep-shallow interface. This is
  better seen in Figure \ref{fig:wavefield-above-step}b, where we plot $\eta$ and $\eta_x$ at
  $(x,y) = (10,0)$ (i.e. above the step) as a function of
  time. We see that, as the drop approaches the wall, the wave height
  above the step (blue curve in Figure
  \ref{fig:wavefield-above-step}b) oscillates in time, indicating the
  presence of a transmitted wave.

The relationship between $\eta$ and $\eta_x$ at the deep-shallow
interface, plotted in Figure \ref{fig:wavefield-above-step}c, appears
to be rather complex. If the variable depth in the fluid were to be
modeled by means of an effective boundary condition of the Robin type,
i.e. $\alpha \eta + \beta \partial \eta/\partial n= \kappa$, then
Figure \ref{fig:wavefield-above-step}c should show (at least
approximately) a straight line;  for the depth ratio
  considered in Figure \ref{fig:wavefield-above-step}, this is not the case. 
Thus, in the context of our model, an effective boundary condition of
the Robin type cannot be used in lieu of the spatially varying
coefficient in the PDE without sacrificing quantitative
  agrement, and a more complex boundary condition must be
sought. Of course, it may very well be that in certain
  limits (e.g. as $h_1 \to 0$) the relation between $\eta$ and
  $\eta_x$ becomes simple, and an effective boundary of the Robin
  type becomes appropriate.

\begin{figure}
    \centering
      \includegraphics[width=1\textwidth]{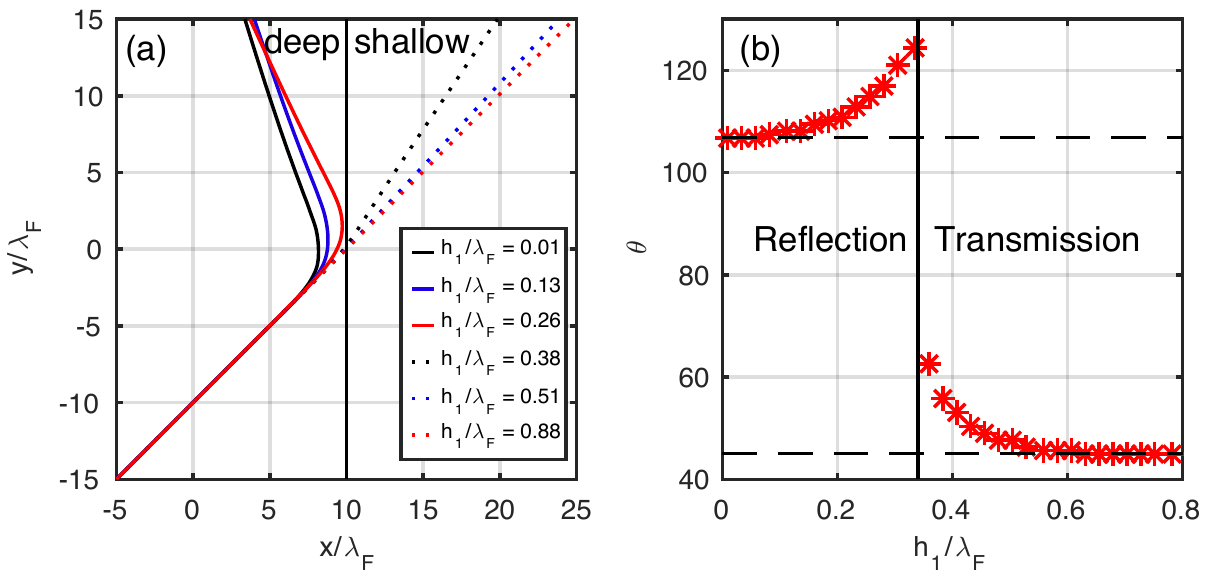}
      \caption{Interaction between a walker and a submerged step as a
        function of the depth above the step. In a) we show a few
        trajectories for depths ranging from very shallow
        ($h_1 \approx 0.01\lambda_F$) to very deep ($h_1 \approx 0.9\lambda_F$). In b) we
        show the final angle, measure relative to the x-axis, as a
        function of the step's depth. The vertical solid line
        separates the drops which were reflected from the drops which
        were transmitted. The dashed horizontal lines indicate the
        asymptotic angles for either very shallow or very deep
        regions. }
\label{fig:reflection-angle-vs-depth}
\end{figure}

We also use the model to investigate the effect of the height above
the step on the trajectory. In Figure
\ref{fig:reflection-angle-vs-depth} we consider a walker coming from a
deep region ($h_0=6.09$mm) and impinging upon a step at an angle of
$45^\circ$. We vary the depth above the step between $0.05$ mm and
$6.09$ mm by small increments, and then calculate the final angle
after the walker has interacted with the step. We see that the final
angle appears to asymptote to a constant for either very shallow or
very deep steps. The drops is unable to feel the bottom
(i.e. the trajectories do not significantly deviate from a straight
line) for depths larger than $\approx 0.6\lambda_F$. For depths less than
$\approx 0.2 \lambda_F$, we also see that the reflection angle depends
only weakly on the depth. This observation, consistent with
experiments (G. Pucci, private communication), suggests that as long as the depth
above the obstacles is small enough, the drop's interaction with boundaries
will be largely independent of the depth in the shallow region
.

\subsection{Diffraction by a slit}\label{sec:diffr-past-slit}

We now briefly consider the case of a walker launched towards a slit. If the
drop behaved like a classical macroscopic particles, we would expect
its trajectory to remain unaffected by the passage through a slit. In
\cite{couder2006single}, it was shown that this is not the case:
passage through a slit deflects the drop owing to the interaction of
its wavefield and the barrier. It was found that walkers behaved (on
average) much like a plane wave diffracted through a slit. Subsequent
experiments conducted by \cite{harris2015pilot},
\cite{andersen2015double}, and \cite{batelaan2016momentum}, were
unable to reproduce the precise diffraction pattern observed in
\cite{couder2006single}.  Nevertheless, walker deflection was
apparent in all of them. \cite{harris2015pilot} found that, in the
absence of external air currents, the single-slit trajectories were
dominated by two preferred angles similar to those emerging in the
reflection experiments. We proceed by showing that our numerical
simulations are consistent with the findings reported in
\cite{harris2015pilot}.

\begin{figure}
    \centering
      \includegraphics[width=1\textwidth]{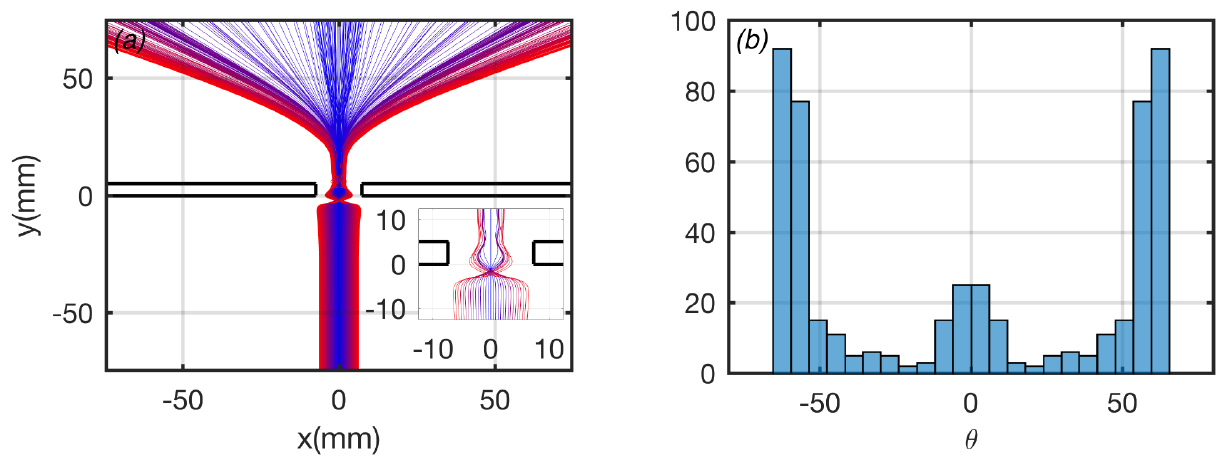}
      \caption{Single slit diffraction. In (a) we show the particle trajectories,
        color-coded according to the impact parameter. In (b) we plot
        the histogram
        of the deflection angle, measured respect to the y-axis. }
\label{fig:slit-diffraction}
\end{figure}

The bath topography consists of a deep region of depth $h_0 = 6.09$mm,
and shallow region, in the shape of a slit, of depth $h_1 = 0.42$mm  (see
figure \ref{fig:slit-diffraction}). The width and breadth of the slit
are respectively $15$mm and $5$mm wide.  Letting $x_i$
be the distance of the initial position of the drop from the center of
the slit, we took a uniform sample of trajectories with impact
parameters ranging from $0$mm to $7.5$mm by increments of $0.025$mm
(300 cases in total). Trajectories with $x_i>6.375$mm reflected from the
barrier, and were thus discarded. All simulations started at
$y=-75$mm, which ensures that a steady walking state is achieved prior
to interaction with the slit. We set $\Gamma/\Gamma_F = 0.99$, and 
considered a drop of radius $0.335$mm, corresponding to the parameters
examined by \cite{harris2015pilot}.


In figure \ref{fig:slit-diffraction}a we show a few sample
trajectories of our numerical simulations. The slit appears to focus
trajectories inside it: its width plays an important role in the
dynamics. The observed pattern does not appear to be of the classical
wave diffraction type; in fact, most trajectories settle to a small
range of angles $\theta$ after passing through the slit, which is
reminiscent of the planar wall reflection and consistent with the
findings reported in \cite{harris2015pilot}. Plotting a histogram of
the final angle of each trajectory (Figure
\ref{fig:slit-diffraction}b), indicates a strong preference for
$55^\circ< \theta < 65^\circ$, which is similar to the preferred angle
observed in \cite{harris2015pilot}. The number of bins in figure
\ref{fig:slit-diffraction}b was chosen to be approximately  $\sqrt{N}$, where $N$ is
the number of trajectories considered. The main qualitative feature of the histogram,
i.e. its lateral peaks, remains unaffected by variations in bin
size. A thorough comparison between our model and recent single and
double-slit experiments will be reported in \cite{pucci2016slit}.



\section{Conclusions}\label{sec:conclusions}

We have presented a model capable of capturing many interesting
features of walker-topography interactions. The model treats depth
changes in the fluid not as boundaries, but as regions where the wave
speed changes, effectively creating a mechanism for partial wave
reflection. Although our equations are reminiscent of a long wave
approximation, we demonstrate that the model remains useful outside
the shallow-water limit provided the system is driven near resonance,
and the drop is synchronized with its Faraday wavefield. 

The assumption that the drop's vertical motion is periodic, and in
resonance with the Faraday waves, might limit the applicability of the
model to resonant walkers. In fact, even drops in the resonant $(2,1)$ modes
may alter their vertical motion, or even violate the periodic assumption, when interacting with other drops
or obstacles. Surprisingly, a reasonable agreement is still
observed in practice (e.g. Figure \ref{fig:wall-reflection-snapshots}), even though we neglect the vertical
dynamics. Inclusion of vertical dynamics into our model is a
possible direction of future research.

Our treatment of the variable topography 
captures the important properties of the reflected waves without a
need to explicitly impose a boundary condition. Although it would be
interesting to develop a systematic understanding of the reflected
Faraday wave, the viscous dissipation and time-dependent gravity are
likely to make a rigorous analysis of the reflection and transmission
laws at a depth discontinuity extremely challenging, even in the
context of our simplified model. Direct numerical simulations of
\eqref{eq:PM-1}--\eqref{eq:PM-4} are also a possible future direction
to better understand the Faraday pilot-waves over variable
topography, although the computational cost of such a task is likely
to be daunting.

The choice of $b(\mathbf{x})$ given by equation
\eqref{eq:effective-depth} was made so as to correctly capture the
phase velocity of the Faraday waves. The approximation of $\phi_z$
given by equation \eqref{eq:wave-eq-approximation}, however, is too
simple to allow for the model to also capture the group velocity of
the Faraday waves correctly. Interestingly, as shown in the comparison
presented in Figure \ref{fig:wavefield-comparison}, the group velocity
of the Faraday waves does not appear to fundamentally affect the
structure of the drop/pilot-wave ensemble. Employing more sophisticated
approximations to $\phi_z$ that also capture the group velocity of
the Faraday waves is likely to improve the quantitative agreement with
the quasi-potential theory of \cite{milewski2015faraday}, and is a
direction currently being pursued.

Finally, we note that all nonlinearities in this hydrodynamic system
arise from the particle-wave coupling, which gives rise to complex
behavior that is neither particle-like nor wave-like. The precise
nature of the coupling is thus extremely import, and cannot be
overlooked in any pilot-wave theory. Although the recent experiments
of \cite{andersen2015double} and \cite{harris2015pilot} do not yield
the specific diffraction pattern reported in \cite{couder2006single},
it is clear that particle-wave duality is at the of heart of the
bouncing drops' quantum-like features. For example, trajectories
deflecting past a slit give clear evidence of the dual nature of bouncing
drops. The relevance of this classical wave-partical duality to quantum
mechanics remains, of course, an open question. Nevertheless, bouncing drops
do provide an enticing physical picture, even if only at the level of analogy, of
what may be happening at scales that we cannot yet probe.

\textbf{Acknowledgements:}
I am grateful to John Bush, Aslan Kasimov, Andre Nachbin, and Ruben
Rosales for fruitful discussions. I would also like to thank Giuseppe Pucci and Pedro Saenz
for making their experimental data available.
\appendix

\section{Details of numerical algorithm}\label{sec:deta-numer-algor}

  In this section we explain how the governing equations are solved
  numerically. The (dimensionless) particle-wave system considered is
  governed by (see section \S \ref{sec:drop-dynamics-wave}):
\begin{align}
\label{eq:appendix-governing-equations-dimensionless-1}
  \phi_{t} &= -G(1 + \Gamma \cos(4\pi  t - \varphi))
                              \eta + \frac{2}{\mathrm{Re}}
                              \Delta \phi+\mathrm{Bo} \Delta
                              \eta -M G \delta(
             {\textbf{x}}-\textbf{x}_p) f(t) \\
  \label{eq:appendix-governing-equations-dimensionless-2}
\eta_{  t} &= -\nabla \cdot ( b \nabla \phi)
+ \frac{2}{\mathrm{Re}}\Delta \eta \\
\label{eq:appendix-EOM-horizontal-dimensionless}
  \frac{d^2{\textbf{x}}_p}{d t^2} &+
  \left(C_{i} f(t) + C_{\text{air}}\right)
  \frac{d{\textbf{x}}_p}{d t} = -G
  f(t)
  \nabla \eta|_{{\textbf{x}} = {\textbf{x}}_p}
\end{align}
where $f(t) =  \sum_{n=0}^\infty \delta(t - n)$. Since the impacts are
instantaneous (i.e. $f(t)$ is a sum of delta functions), it is
convenient to divide the time evolution in two parts: impact and
free-flight. The algorithm consists of two main routines: one which
evolves the particles and the waves during free-flight, and another
which resolves the details of the impact. 

During free-flight the wave evolution is solved using a (Fourier)
pseudo-spectral method in space. Taking the Fourier transform of
\eqref{eq:appendix-governing-equations-dimensionless-1}--\eqref{eq:appendix-governing-equations-dimensionless-2},
and assuming $t\neq 1,2,...$ so that $f=0$, 
yields
\begin{align}
\label{eq:appendix-governing-equations-dimensionless-Fourier-1}
  \hat{\phi}_{t} &= -G(t) \hat{\eta} - \frac{2 k^2}{\mathrm{Re}}
                              \hat{\phi} - \mathrm{Bo}k^2 \hat{\eta}\\
  \label{eq:appendix-governing-equations-dimensionless-Fourier-2}
\hat{\eta}_{  t} &= -i \mathbf{k} \cdot \widehat{( b \nabla \phi)}
             - \frac{2 k^2}{\mathrm{Re}} \hat{\eta} 
\end{align}
The time evolution of the waves is carried in Fourier space by a
standard fourth order Runge-Kutta method
(e.g. \cite{leveque2007finite}). The topography enters through the term $\widehat{(b \nabla
  \phi)}$, which is numerically computed pseudo-spectrally in the
following way:
\begin{align}
  \label{eq:pseudospectral-term}
  \mathcal{F}[b(\mathbf{x}) \nabla
  \phi] &= \mathcal{F}\left[ b(\mathbf{x}) \mathcal{F}^{-1}[i \mathbf{k} \mathcal{F}[\phi]] \right]
\end{align}
For the topography considered in the paper, $b(\mathbf{x})$ is a
discontinuous function, and the Fourier transform in
\eqref{eq:pseudospectral-term} will be subject to the well known Gibbs
phenomenon. Although in principle this could generate numerical
stability problems and/or degrade the accuracy of the algorithm, we
observed that in practice the damping provided by the viscous terms in
\eqref{eq:appendix-governing-equations-dimensionless-1}--\eqref{eq:appendix-governing-equations-dimensionless-2}
was enough to smooth the solution sufficiently so that no numerical
instabilities were observed (see also Appendix \ref{sec:conv-tests-numer} for
convergence tests). Similarly, during free-flight the particle
evolves according to
$\ddot{\mathbf{x}}_p + C_{\text{air}}\dot{\mathbf{x}}_p = 0$, which is
solved analytically.

During impact, we exploit the temporal delta function present in the
model to resolve the interaction analytically. That is, at impact
times $t_i = {1,2,...}$ we modify $\phi$ and $\mathbf{x}_p$ by
integrating our equations from immediately before until immediately
after an impact. For the surface potential $\phi$, integration of
\eqref{eq:appendix-governing-equations-dimensionless-Fourier-1} across
an impact can be easily shown to yield:
\begin{align}
  \label{eq:phi-updated-upon-impact}
  \mathcal{F}[\phi^+] &= \mathcal{F}[\phi^-] + MG
                        \mathcal{F}[\delta(\mathbf{x} - \mathbf{x}_p)]
                        = \mathcal{F}[\phi^-] + M G e^{i
                        \mathbf{k}\cdot \mathbf{x}_p},
\end{align}
where the plus and minus superscripts denote values immediately after
and immediately before the impact, respectively. Integration of
\eqref{eq:appendix-EOM-horizontal-dimensionless} requires a bit more
care since the friction term contains a delta function, which
multiplies $\dot{\mathbf{x}}_p$, a discontinuous term. For simplicity
we consider the case of a single impact at $t = t_i$. Immediately
before the impact both $\eta$ and $\mathbf{x}_p$ are known, and
continuity of both across the impact implies
$\eta(\mathbf{x},t_i^-) = \eta(\mathbf{x},t_i^+)$ and
$\mathbf{x}_p(t_i^-) = \mathbf{x}_p(t_i^+)$. For convenience, we define
$\mathbf{\alpha} = -G \nabla \eta(\mathbf{x}_p(t_i),t_i)$. Resolving
\eqref{eq:appendix-EOM-horizontal-dimensionless} across an impact is
then equivalent to solving
\begin{align}
  \label{eq:delta-ode}
    \ddot{\textbf{x}}_p&+
  C_{i} \delta(t-t_i)
  \dot{\textbf{x}}_p = \mathbf{\alpha} \delta(t-t_i).
\end{align}
We proceed formally by multiplying \eqref{eq:delta-ode} by the
integrating factor $I = e^{C_{i}{\int \delta(t-t_i) dt}} = e^{C_i
  H(t-t_i)}$, yielding:
\begin{align}
  \label{eq:delta-ode-integrated}
    \frac{d}{dt}\left( I(t) \dot{\mathbf{x}}_p \right) &= \mathbf{\alpha} I(t) \delta(t-t_i),
\end{align}
and so
\begin{align}
  \label{eq:7}
      e^{C_i} \dot{\mathbf{x}}_p^+ - \dot{\mathbf{x}}_p^-  &= \mathbf{\alpha}
                                             \int_{t^-}^{t^+} e^{C_iH(t-t_i)}
                                             \delta(t - t_i) dt =
                                             \frac{\mathbf{\alpha}}{C_i} (e^{C_i}-1).
\end{align}
Solving for $\dot{\mathbf{x}}_p(t_i^+)$ gives:
\begin{align}
  \label{eq:3}
  \dot{\mathbf{x}}_p^+ &= \dot{\mathbf{x}}_p^-e^{-C_i}
                         -
                         \frac{G}{C_i}(1-e^{-C_i})\nabla \eta|_{\mathbf{x}=\mathbf{x}_p},                       
\end{align}
which is used in the algorithm to update the drop's velocity at each
impact\footnote{A more rigorous way to proceed is to
  consider an $\epsilon$ mollification of the Dirac delta, solve
  \eqref{eq:delta-ode} with $\delta$ replaced by the
  $\delta_\epsilon$ and then take the limit $\epsilon \to 0$ at the
  end. The final result of this procedure is the same as that
  obtained from our formal calculation.}.


\section{Convergence tests}\label{sec:conv-tests-numer}

\begin{figure}
    \centering
      \includegraphics[width=1\textwidth]{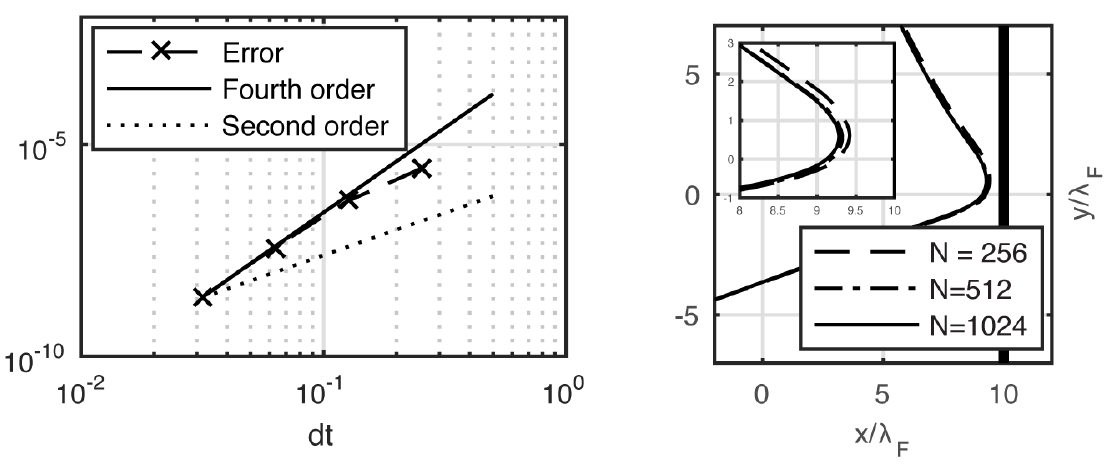}
      \caption{Convergence tests. In (a) we show fourth order  (self)
        convergence of the algorithm in freespace. The error shown in
        (a) is defined as maximum error (over the last 50
        impacts) between the numerically observed walking speed to the
        walking speed of a very fine simulation, used as a
        reference. As expected, fourth order convergence in time is
        recovered. In (b) we show the effect of spatial
        resolution on drop's trajectory in the presence of a depth
        discontinuity (located at $x=10$). The domain size is 64x64, corresponding to the
        results presented in \S \ref{sec:refl-from-plan}. Since
        trajectories with resolution of 512x512 modes were very
        similar to those with 1024x1024, the 512x512 resolution was
        adopted throughout this paper and considered enough for
        reliable results.}
\label{fig:convergence-tests}
\end{figure}

In this section we investigate the effect of the numerical resolution
on the results obtained. First we verify that the walking speed in
freespace is not a function of the resolution, and that as the time
steps are decreased the algorithm converges with fourth order accuracy. This is
shown in Figure \ref{fig:convergence-tests}a, where we plot the error in
the free-walking speed as a function of $\Delta t$ for a fixed spatial
resolution of 512x512 modes. Each simulation was ran until the
walker's speed saturated to a constant, and the maximum error
(relative to a very fine simulation) over the
last 50 impacts was computed. Clearly, despite
our treating the drop as a delta function in space and time, fourth order
convergence is still recovered. 

In Figure \ref{fig:convergence-tests} we verify that the
resolution of 512x512 spectral modes in a domain of size 64x64, which
was used in the simulations presented in \S
\ref{sec:numer-simul-comp}, is sufficient to obtain a reliable
result even in the presence of a piecewise constant topography. More
precisely, we plot the trajectory of a walker undergoing reflection using 256,
512, and 1024 spectral modes in each direction. As can be seen in
Figure \ref{fig:convergence-tests}b, the difference between
using 512 and 1024 modes was sufficiently small that 512 was considered an
acceptable resolution (in fact the final reflection angles predicted using 512 or
1024 modes vary by
less than $1^\circ$). 

\clearpage
\bibliographystyle{plain}
\bibliography{lfaria-refs}

\end{document}